\documentclass[12pt,a4paper]{article}
\usepackage[unicode]{hyperref}

\usepackage{graphicx}
\usepackage{amssymb,amsmath}
\usepackage[cp1251]{inputenc}  
\usepackage[T2A]{fontenc}
\usepackage[russian,english]{babel}
\usepackage{caption}

\usepackage{indentfirst}

\title{\bf THE ``RELATIVISTIC'' MUG}
\author{L.~B.~Okun\\
ITEP, Moscow, Russia}
\begin{document}
\date{20.10.10}

\maketitle

\bigskip

\begin{abstract}

This note is an attempt to explain in simple words why the famous relation 
$E=mc^2$ misrepresents the essence of Einstein's relativity theory. The  note is 
addressed to high-school teachers, and a part of it -- to those university 
professors who permit themselves to say that the mass of a body increases with 
its velocity or momentum and thus mislead the teachers and their students. \end{abstract}

\section{Introduction}
 
The moral health of the modern society and its material well-being are unthinkable without high status of science in the country. This status in a certain degree depends on how adequate is the image of science in the mirror of mass culture.
For more than twenty years I have been collecting artifacts of mass culture (from postcards and T-shirts to popular articles and books) sporting ``the famous Einstein's formula''. 
 
Recently my friends added to my collection a Relativity Floxy Noxy mug. You can 
google (typing these four words in the search line of your computer) and see it:

\begin{figure}[h]
	\begin{center}
	\scalebox{1} {\includegraphics{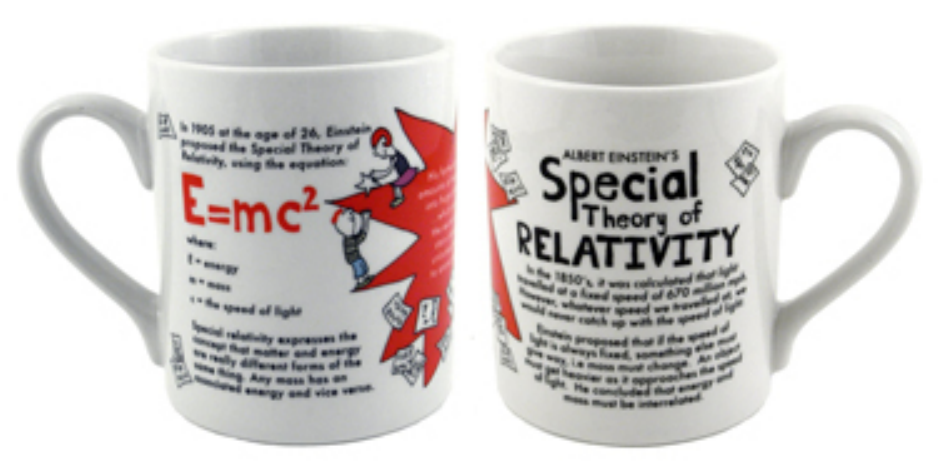}}
\end{center}
\caption*{The Relativity Mug}
\end{figure}

In a certain sense it contains the quintessence of my collection presenting the 
main popular science clich\`es and misconceptions. As they are quite often 
repeated in newspapers and textbooks, I decided to reproduce the text on the mug 
and to explain briefly what is wrong with it. I believe that it may be useful 
to many people.

\section{The text on the mug}
There are three columns of text on the mug --  to the right of the handle ({\bf 1}), to the left of the handle ({\bf 2}),  opposite the handle ({\bf 3}):

{\bf Column 1.} In 1905 at the age of 26, Einstein 

proposed the Special Theory 

of Relativity, using the equation: 

$E=mc^2$ 

where 

$E$=energy, 

$m$=mass, 

$c$=the speed of light.

Special relativity expresses the 

concept that matter and energy 

are really different forms of the 

same thing. Any mass has 

an associated energy and vice versa.

{\bf Column 2.} Albert Einstein's 

SPECIAL 

Theory of 

RELATIVITY

In the 1850's it was calculated that light 

traveled at a fixed speed of 670 million mph. 

However, whatever speed we traveled at, we 

would never catch up with the speed of light.

Einstein proposed that if the speed of 

light is always fixed, something else must 

give way, i. e. mass must change. An object 

must get heavier as it approaches the speed 

of light. He concluded that energy and 

mass must be interrelated.

{\bf Column 3.} His formula suggested that tiny 

amounts of mass  can be converted 

into huge amounts of energy... 

... which revealed the secret of how 

stars shine and 

unlocked the key 

to atomic energy.

\section{My clarifications and comments}

{\bf Column 1.} As is well known, Maxwell in 1860-70's united optics with electricity 
and magnetism by establishing equations describing not only static fields, but 
also alternating electromagnetic fields propagating in a vacuum with velocity of 
light. Several physicists in the 1880-90's after realizing that Maxwell 
equations are incompatible with equations of Newton mechanics have attempted to 
preserve the latter at velocities comparable to that of light by postulating 
that all new physics can be reduced to the increase of mass of a body with its 
velocity. These attempts were continued in the 20th century; they are briefly 
described in the articles \cite{lo3}, \cite{lo4}. However it became clear rather 
soon that some of the Newton equations cannot be preserved, for instance the 
famous equation ${\bf F}=m{\bf a}$, where ${\bf F}$ is force and ${\bf a}$ is acceleration. It turned out that the equations themselves should be 
changed in such a way that mass would not depend on velocity, but there would 
exist an important new link between mass and energy. Let us note that relation between force and momentum remained the same: ${\bf F}=d{\bf p}/dt$, but, as we will see below, the relation between momentum and velocity ${\bf p}=m{\bf v}$ has changed. This had serious impact on 
the language and philosophy  of physics.   

In summer of 1905 Einstein published a detailed article \cite{ein1} in which he 
presented his theory which later got the name Einstein's theory of relativity. 
This theory extended to electromagnetic phenomena the principle of relativity 
formulated by Galileo and Newton. According to it, it is impossible by any 
experiment to find out whether a closed space (say, a cabin of a ship) is at 
rest or in a uniform and rectilinear motion. Soon the theory was extended to the 
newly discovered nuclear phenomena and got the name Special Relativity (SR). 
This theory describes the motion and interaction of fast particles whose 
velocities are comparable with the speed of light. Such particles are 
called relativistic. (In 1915 Einstein proposed General Relativity (GR) to 
describe gravity. But in this note we will not consider it.)

In autumn of 1905 Einstein published a short note \cite{ein2} in which he stated that in the framework of his theory the mass of a body is a measure of its energy content. 
The total energy $E$ of a free  body is equal, according to the theory of relativity, to the sum of its kinetic energy $E_K$ (of the motion as a whole) and its energy at rest -- rest energy $E_0$:
\begin{equation}
E=E_K + E_0. 
\end{equation}
Of course, the concept of a free (isolated from any external influence) body is an idealization. But idealization (abstraction) lies at the basis of scientific method and is extremely fruitful. 

The realization that any body at rest possesses energy was the greatest discovery of the 20th century. The amount of this energy is given by Einstein's equation: 
\begin{equation}
E_0=mc^2, 
\end{equation}
where $m$ is the mass of the body and $c$ is the speed of light. (It was exactly in this form that Einstein had written  equation (44) in 1921  in his lectures ``The meaning of relativity'' \cite{ein3}, though the notion of the rest energy $E_0$ appeared already in the note \cite{ein2}.)

The kinetic energy of ordinary bodies is given by the well known equation of Newton's mechanics $E_K=mv^2/2$. As the velocity $v$ of an ordinary body is much less than $c$, the rest energy of a body is huge in comparison with its kinetic energy. But in the ordinary life the rest energy does not manifest itself. Einstein pointed out that part of it is liberated in the radioactive decays.

Unfortunately, many famous physicists during the last century have formulated the Einstein equation in a ``simplified form'' by omitting the index zero:
\begin{equation}
E=mc^2,
\end{equation}
and treating this relation as increase of mass not only with energy but also with velocity and momentum of the body. 

In 1948 Einstein warned  Barnett -- the author of the book ``Universe of Dr. 
Einstein'' -- against using the concept of mass depending on velocity. (A copy 
of this handwritten letter is reproduced in ref \cite{lo3}.) But sometimes, 
especially in his popular writings, he himself did not care about the index zero. 
This semantic kink was caused by the clash of two languages -- the old 
non-relativistic and the new, consistently relativistic one.   
 
{\bf Column 2.} The assertion that the speed of light is always fixed at a value of 670 
million mph is correct, but the dating (1850's) is not quite correct. That the 
speed of light is finite (not infinite), was established in 1676 by R\"{o}mer 
who deduced this from observations of Jupiter's satellite. It followed from them 
that the speed is around 200 000 km/s. The first and more precise measurements 
of $c$  on the Earth were performed by Fizeau in 1849. But the fact that the 
speed about 300 000 km/s is fixed and does not depend on the velocities of the 
source and the observer,  was discovered in 1887 by Michelson and Morely.

The statement that energy and mass are interrelated is correct: $E_0= mc^2$, 
while that the mass changes with velocity is definitely wrong. In the theory of 
relativity (unlike the mechanics of Newton) the measure of inertia is not mass 
$m$ but the total energy $E$ of the body. The momentum $\bf p$ of a body is 
connected with its velocity $\bf v$ not by the Newton's relation ${\bf p}=m{\bf 
v}$ but by the relation 
\begin{equation}
{\bf p}=(E/c^2){\bf v}. 
\end{equation}
As a result it is the more difficult to change the momentum of a body, the higher its total energy $E$. And $E/c^2=m$ only at zero momentum, when the total energy equals the rest energy $E_0$. 

One can feel more deeply that the measure of inertia is energy by considering 
the example of the Large Electron-Positron collider LEP which operated at CERN 
during the last decade of the 20th century. Particles with energy 50 GeV were 
kept in its 27 km ring tunnel by a rather weak field of iron magnets. (Without 
this field particles would fly along a strait line.) Exactly the same field 
would maintain the circulation of protons with the same momentum (and almost the 
same energy), though the mass of the proton is 2000 times larger than the mass 
of the electron. In the year 2010 the Large Hadron Collider LHC started to 
operate in the same tunnel. To circulate protons with energy 3500 GeV, the 
magnetic field of superconducting magnets in it is 70 times stronger. 

Thus, the measure of inertia of a particle is its 
total energy 

{\bf Column 3.} Here everything is correct if one uses the equation $E_0=mc^2$ and takes into account that in nuclear reactions in the stars, in the Sun and on the Earth a part of the rest energy of the particles which are burned is transformed into kinetic energy of the products of burning. The same is valid for any process of burning.

\section{Four dimensions of the world}

Now I would like to address a few words to those who are more or less familiar with 
the concept of four-dimensional world (4-world) introduced in the relativity 
theory in 1908 by Minkowski \cite{min}. In the 4-world the time coordinate $ct$ 
of an event and its position coordinates $\bf r$ form a 4-vector. Similarly the 
energy $E$ of a free (isolated) body (more precisely, $E/c^2$) and three 
components of its momentum $\bf p$ (more precisely, ${\bf p}/c$ ) form four 
components of the pseudo-euclidean 4-vector.  The scalar length of this 4-vector 
is given by the mass of the body $m$ according to the equation \begin{equation} 
m^2=E^2c^{-4}-{\bf p}^2c^{-2}. \end{equation} (The words ``pseudo euclidean'' 
indicate that the square of the length of the 4-vector is equal not to the sum 
but to the difference of squares of its $E$- and $\bf p$-components.)

Taylor and Wheeler in the book \cite{tw} put energy and momentum on the 
orthogonal axes, then on the hypotenuse they depict mass by a short and thick 
segment. But it is possible to present equation (5) simply as a right triangle 
if one  rewrites it in the form $E^2=m^2c^4+{\bf p}^2c^2$  and puts 
mass and momentum on the orthogonal axes (see article \cite{lo5}). Then  energy 
is the hypotenuse, while mass and momentum are the other two legs. For  any value of 
momentum the kinetic energy is
\begin{equation}
E_K=\sqrt{m^2c^4+{\bf p}^2c^2}-mc^2.
\end{equation}

The main equation (5) of relativity theory has been tested in thousands of 
experiments with the accuracy of up to ten digits. For a massive body whose 
momentum is zero it implies $E_0=mc^2$. For a non-vanishing momentum one can 
rewrite it as $(Ec^{-2}-m)(Ec^{-2}+m)={\bf p}^2c^{-2}$ and at $E_K \ll E_0$ 
derive from it the non-relativistic expression for kinetic energy $E_K={\bf 
p}^2/2m$ without developing the square root. Similarly, for relativistic particles $E-|{\bf p}|c=mc^2/2E$. (This equation is essential for neutrino oscillations.) It follows also from equation (5) and from the formula (4) for velocity ${\bf v}={\bf p}c^2/E$ that for a 
massless particle of light -- the photon -- the speed is always equal to $c$. 

The special theory of relativity is impeccable. One cannot say the same about its image in the mass culture. 

Unfortunately, the sudden illness and death of Minkowski did not allow him to persuade his contemporaries to switch to the language of the four-dimensional world, and they continued futile attempts to explain the meaning of relativity theory in terms of Newton's three-dimensional mechanics. Though Einstein used the four-dimensional mathematical apparatus in deriving the equations of his general theory of relativity for gravitational interaction, I failed to find the equation $E^2c^{-4}-{\bf p}^2c^{-2}=m^2$ on the pages of his writings. 

It appeared first  in the articles of Klein \cite{kle}, Fock\cite{fock}, Gordon\cite{gor} (1926) and especially in the works of Dirac \cite{dir} (1930) in which relativistic quantum mechanics was constructed (as is well known, Einstein, a co-founder of the concept of quantum, did not accept quantum mechanics). The equation appeared in the framework of not quantum but classical field theory much later, in the book ``The classical theory of fields'' by Landau and Lifshitz in 1941 (in Russian) \cite{ll}.  

Four-dimensional description is equally good for massive and massless particles of matter. It shows that mass and matter are not the same thing, that energy and momentum are the measures of all processes and motions in nature. As for the mass of the particles, it becomes non-essential for processes at high energies $E \gg mc^2$.  

\section{The speed of light as the unit of velocity}
The correct equations must be correct regardless of the choice of units. The existence 
of the universal maximal velocity $c$ allows one to express any velocity $v$ in 
units of $c$ as a dimensionless number $\beta=v/c$. It is evident that in these 
units $\beta=1$ for $v=c$. As a result one can get rid of $c$ in the equations 
of relativity theory by rewriting equations (2), (4), (6) in the form
\begin{align}
E_0=m,\hspace{7mm}
m^2=E^2-{\bf p}^2,\hspace{7mm}
{\bf v}={\bf p}/E.
\end{align}
As for equation (3) $E=mc^2$, it is reduced to $E=m$, which evidently contradicts equation (1) $E=E_K+E_0=E_K+m$ and hence is wrong. 

\section{Conclusion} 
Volodya Gribov, whose attitude concerning $E=mc^2$ was the same as mine, 
gave me a friendly advice in 1980s not to struggle against the famous and false equation because 
this fight just cannot be won. It was with a feeling of permanent defeat that I was 
writing the text above for the forthcoming volume ``Gribov-80'' in the summer 
2010, amidst the unprecedented heat and smog in Moscow, which could 
but adversely affect the quality of the text. In autumn, a few weeks ago, Julia 
Nyiri reminded me that this text is a continuation of my contribution to the 
volume ``Gribov-75'' \cite{lo7} in  which I compared the equation $E=mc^2$ with
a virus. Indeed, the concept of relativistic mass hidden in the equation 
$E=mc^2$ is a semantic virus similar to computer viruses. People infected by 
this virus (they often call themselves relativists) believe that Relativistic 
Mass is the main portal to Relativity Theory because mass is the measure of 
inertia. They ignore the fact that mass is the measure of inertia only for very 
slowly moving bodies and particles for which the rest energy $E_0$ is much 
larger than the kinetic energy $E_K$. When velocities are not very low, mass is 
only an approximate measure of inertia. For fast particles for which $E_K \gg 
E_0$ (photons, neutrinos, protons in LHC) the measure of inertia is the total 
energy $E$.  

It is well known that formulas in physics are a continuation of the ordinary 
language: equations are encoded sentences, while mathematical symbols in these 
equations are encoded words or terms. To prevent confusion, each symbol must be 
unambiguously connected with a corresponding term. Is it possible to introduce 
in Special Relativity the concept of relativistic mass? Yes, it is possible: 
$m_r=E/c^2$, though it will be just another symbol for energy because $c$ is a 
universal constant. Is it possible then to introduce the term relativistic rest 
mass $m_{r0}=m$? Yes, it is possible. Of course, it is equally possible to introduce 
both terms and both symbols. Although they are not needed in Special Relativity 
as it is a complete, self consistent theory without them, their 
introduction is possible. What is not good is to denote the relativistic rest 
mass $m_{r0}$ by $m_0$ and then call it simply rest mass, because this presumes 
that mass $m$ depends on velocity in Special Relativity (SR), while we all know 
that in SR $m$ is Lorentz invariant: it is the same at rest and in motion, and 
hence, there is no sense in supplying it with indices.  

But the real trouble begins when $m_r$ is called the mass, is denoted by $m$ and 
at the same time the ordinary Newtonian mass $m$ is renamed into $m_0$. Then the 
mixing of two languages (``French and Nizhegorodsky'') mutilates the beautiful 
theory, leads to unbelievable confusion and thwarts its understanding. To top it 
all, some `` philosophers-relativists'' then allege that the mechanics of Newton 
is not a limiting case of mechanics of Einstein, and that these two theories are 
incommensurate. 

As a result of the reverse action of mass culture on the scientific culture, 
many chapters in the best text-book on physics of the 20th century published in 
1960s -- ``The Feynman Lectures on Physics''\cite{fey} -- repeat the statement 
that mass changes with velocity. The little book ``What is 
relativity?''\cite{lr} by Landau and Rumer also claims that mass increases with 
velocity. (The book was written in the 1930s, before the arrest of both authors, 
and published in the 1950s after Rumer was released from exile. (Landau remained in 
jail for one year.)) ``The classical theory of fields'' by Landau and Lifshitz 
published in 1940s was the first text-book in the world in which mass was 
velocity-independent. But even in it the concept of rest energy $E_0$ was missing
and the Einstein's formula was mentioned in the form $E=mc^2$. This discrepancy
is kept in the latest edition of the book in the 21st century.Indeed, nobody is 
perfect. Our language is not perfect: ``a spoken thought is a lie''.

It is impossible in this short note to refer to the articles and books of the 
creators of relativity theory, but it is easy to find them by clicking the hyper 
references \cite{lo3}, \cite{lo4}, \cite {lo5},\cite{lo7}, \cite{lo1}, 
\cite{lo2} listed below. The seventh hyper reference \cite{lo6} contains slides of the talk 
 which explained why the teaching of physics must be based on two 
fundamental constants of nature: $c$ and $\hbar$. By operating with these two 
constants I plan to present the foundations of physics in a little book (100 
pages) ``The ABC of Physics''.

\vspace{3mm}

{\bf Acknowledgments}

\vspace{3mm}

I am grateful to Erica Gulyaeva, Marek Karliner, Elya and Vitaly Kisin, Olga Milyaeva, Boris Okun and Zurab Silagadze whose remarks helped me write this note.

The work is supported by grant of the President of RF  NSh-4172.2010.2

\end{document}


\date{24.10.10}

\maketitle

\bigskip

\begin{abstract}
Эта заметка представляет собой попытку просто объяснить, почему знаменитое соотношение $E=mc^2$ неправильно выражает суть теории относительности Эйнштейна.  Заметка в основном адресована школьным преподавателям, а часть её -- тем университетским профессорам, которые позволяют себе говорить, что масса тела растёт с ростом его скорости или импульса, и тем самым вводят в заблуждение школьных преподавателей и их учеников.

\end{abstract}

\section{Введение}

Нравственное здоровье современного общества и его материальное благополучие невозможны без высокого статуса науки в стране. 
А это, в известной мере определяется тем, насколько адекватно наука отражается в зеркале массовой культуры. 
Более двадцати лет я собираю артефакты массовой культуры (от открыток и футболок до научно-популярных статей и книг), посвящённые "`знаменитой формуле Эйнштейна"'.

Недавно друзья подарили мне кружку, которая называется Relativity Floxy Noxy mug. Если набрать эти четыре слова в поисковике компьютера, то можно рассмотреть кружку.
\begin{figure}
	\begin{center}
	\scalebox{1} {\includegraphics{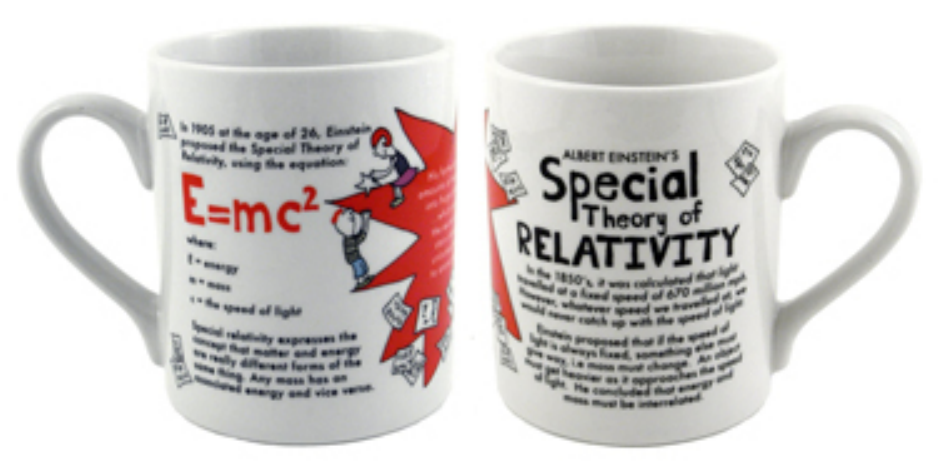}}
\end{center}
\caption*{Кружка Relativity}
\end{figure}

Она в некотором смысле представляет собой квинтэссенцию моей коллекции, так как содержит основные научно-популярные штампы по этой теме. Поскольку эти штампы продолжают попадать в газеты и в учебники, я решил воспроизвести надписи на чашке и кратко объяснить, что в них неверно. Думаю, что получившаяся заметка может быть полезна многим. 
  
\section{Перевод на русский язык текста на кружке}

На кружке имеются три колонки текста --  направо от ручки ({\bf 1}), налево от ручки ({\bf 2}), напротив ручки ({\bf 3}):

{\bf Колонка 1.} 

В 1905 году 26-летний Эйнштейн

предложил Специальную Теорию 

Относительности, используя уравнение: 

$E=mc^2$, 

где 

$E$=энергия,

$m$=масса, 

$c$=скорость света.

Специальная теория относительности выражает 

мысль, что материя и энергия 

являются в сущности различными формами 

одного и того же. Любая масса имеет 

связанную с ней энергию, и наоборот.

{\bf Колонка 2.} 

СПЕЦИАЛЬНАЯ 

теория 

ОТНОСИТЕЛЬНОСТИ 

Альберта Эйнштейна.

В 1850-х годах было вычислено, что свет

движется с фиксированной скоростью 

670 миллионов миль в час. 

Однако, как бы быстро мы ни двигались, 

догнать свет мы не можем.

Эйнштейн предложил, что если скорость света 

всегда фиксирована, то что-то другое должно

изменяться, а именно должна изменяться масса. Предмет 

должен становиться тяжелее по мере того, как его скорость приближается 

к скорости света. Он заключил, что энергия и 

масса должны быть связанными друг с другом.

{\bf Колонка 3.} 

Его формула означает, что маленькие 

количества массы можно превратить 

в огромные количества энергии... 

...что раскрыло секрет того, как 

сияют звёзды, и 

послужило ключом 

к атомной энергии.

\section{Мои уточнения и комментарии}

{\bf Колонка 1.} Как известно, в 1860-70-х годах Максвелл объединил оптику с физикой электрических и магнитных явлений, открыв уравнения, описывающие не только статическое, но и движущееся в вакууме со скоростью света электромагнитное поле. В 1880-90-х годах ряд физиков, обнаружив, что уравнения Максвелла не согласуются с уравнениями механики Ньютона, попытались сохранить уравнения Ньютона и при скоростях, сравнимых со скоростью света, запрятав всё новое в гипотезу о том, что масса тела растёт с ростом его скорости. Эти попытки продолжались и в ХХ веке; они кратко описаны в статьях \cite{lo3}, \cite{lo4}. Однако довольно скоро выяснилось, что так можно поступить не со всеми уравнениями   Ньютона; например, так нельзя поступить со знаменитым уравнением ${\bf F}=m{\bf a}$, где {\bf F} --сила, а  {\bf a} -- ускорение. Оказалось, что надо изменить сами уравнения механики Ньютона, сохранив при этом независимость массы от скорости, но установив новую, очень важную связь массы и энергии.  Заметим, что связь силы с импульсом при этом сохранилась: ${\bf F}=d{\bf p}/dt$, но связь импульса cо скоростью ${\bf p}=m{\bf v}$ стала, как мы увидим, иной. Всё это сопровождалось изменением языка и философии физики. 

Летом 1905 года Эйнштейн опубликовал детальную статью \cite{ein1}, в которой предложил теорию, вскоре названую теорией относительности Эйнштейна, поскольку в ней на электромагнитные явления был распространён принцип относительности Галилея-Ньютона. Согласно этому принципу, никакими опытами внутри замкнутого пространства (например, каюты корабля) невозможно выяснить, движется ли оно прямолинейно и равномерно, или покоится. Затем эту теорию распространили и на только что открытые ядерные явления, и за ней закрепилось название Специальная Теория Относительности (СТО). По существу, СТО -- это теория движения и взаимодействия быстрых частиц, скорость которых сравнима со скоростью света. Такие частицы стали называть релятивистскими. (В 1915 году Эйнштейн предложил Общую Теорию Относительности (ОТО), описывающую гравитационные явления. Но в этой заметке мы её касаться не будем.)

Осенью 1905 года появилась короткая заметка Эйнштейна \cite{ein2} о том, что масса тела есть мера содержащейся в нём энергии. В теории относительности полная энергия свободного  тела $E$ равна сумме его кинетической энергии, т.е. энергии движения тела как целого $E_K$ и энергии, содержащейся в покоящемся теле, т.е. энергии покоя $E_0$:
\begin{equation}
E=E_K + E_0. 
\end{equation}
Разумеется, представление о свободном теле, изолированном от всех внешних воздействий, является идеализацией, абстракцией.
Но абстракция лежит в основе научного метода и является исключительно плодотворной.
 
Осознание того, что любое покоящееся тело обладает энергией, было величайшим открытием ХХ века. Величина этой энергии выражается через массу тела $m$ соотношением Эйнштейна: 
\begin{equation}
E_0=mc^2. 
\end{equation}
(Именно в таком виде написал Эйнштейн уравнение (44) в лекциях "`Сущность теории относительности"' \cite{ein3}, прочитанных им в 1921 году, хотя понятие энергии покоя $E_0$ появилось уже в заметке \cite{ein2} 1905 года.) 

Поскольку кинетическая энергия обычных окружающих нас тел даётся известной формулой механики Ньютона $E_K=mv^2/2$ и поскольку скорость обычных тел $v$ много меньше скорости света $c$, то их кинетическая энергия неизмеримо меньше их энергии покоя. Но в каждодневной жизни энергия покоя не проявляется. Эйнштейн сразу же заметил, что часть этой энергии выделяется при распаде радиоактивных ядер.

К сожалению, как в начале, так и на протяжении всего ХХ века, многие известные физики формулировали соотношение Эйнштейна в "`упрощённом виде"', опуская индекс ноль у энергии: 
\begin{equation}
E=mc^2
\end{equation}
и трактуя это соотношение как возрастание массы с ростом не только энергии, но и импульса, и скорости тела. 

Известно письмо 1948 года Барнетту --- автору книги "`Вселенная доктора Эйнштейна"', в котором Эйнштейн предупреждает о том, что не следует пользоваться понятием массы, зависящей от скорости (смотрите копию рукописи этого письма в \cite{lo3}). Но временами, особенно в научно-популярных текстах, и он опускал индекс ноль.
По существу, семантическое завихрение в его творчестве возникло при столкновении двух языков -- старого нерелятивистского и нового релятивистского.

{\bf Колонка 2.} Утверждение о том, что свет всегда имеет фиксированную скорость 670 миллионов миль в час, -- правильное, но утверждение, что это было вычислено в 1850-х годах, не вполне верное. Что скорость света конечна, а не бесконечна, выяснил в 1676 году Рёмер, определивший её по наблюдениям спутника Юпитера. Из них следовало,что она составляет примерно 200 000 км/с. Первые и более точные измерения в земных условиях произвёл Физо в 1849 году. А то, что скорость 300 000 км/с с высокой точностью фиксирована, т.е. не зависит от скорости источника света и наблюдателя, обнаружили в 1887 году Майкельсон и Морли. 

Заключение о том, что энергия и масса связаны друг с другом, несомненно правильное, поскольку $E_0=mc^2$. А вот утверждение о том, что масса тела должна меняться вместе с его скоростью, заведомо неверное. В теории относительности, в отличие от механики Ньютона, мерой инерции является не масса тела $m$, а его полная энергия $E$, а импульс тела $\bf p$ связан с его скоростью $\bf v$ не соотношением Ньютона ${\bf p}=m{\bf v}$, а соотношением 
\begin{equation}
{\bf p}=(E/c^2){\bf v}. 
\end{equation}
В результате изменить импульс тела тем трудней, чем больше его энергия $E$. И только при нулевом импульсе, когда $E=E_0$, т.е. когда полная энергия равна энергии покоя, только тогда $E/c^2=m$ .  

Прочувствовать, что именно энергия частицы является мерой её инерции можно на примере Большого Электрон-Позитронного коллайдера LEP, работавшего в ЦЕРН в последнее десятилетие ХХ века. В его кольцевом туннеле длиной 27 км электроны и позитроны с энергией около 50 ГэВ летели навстречу друг другу. Удерживались они в этом кольце довольно слабым  полем железных магнитов, расположенных в туннеле. (Если бы не было поля, частицы летели бы по инерции по прямой линии.) 
Точно так же удерживало бы это поле и протоны с  тем же импульсом и почти той же энергией, несмотря на то, что масса протона почти в 2000 раз больше массы электрона. В 2010 году в том же туннеле заработал Большой Адронный Коллайдер LHC. Чтобы удерживать в нём протоны с энергией 3500 ГэВ, понадобилось в 70 раз более сильное поле сверхпроводящих магнитов. 

Итак, мерой инерции частицы является её полная энергия.
 
{\bf Колонка 3.} Здесь всё правильно, если говорить о формуле $E_0=mc^2$ и учесть, что в ядерных реакциях в звёздах, на Солнце и на Земле часть энергии покоя сгорающих частиц превращается в энергию движения продуктов горения. Это же, разумеется, относится и ко всем процессам горения. 

\section{Четыре измерения мира}
Теперь я хотел бы сказать несколько слов тем, кто более или менее знаком с четырёхмерным математическим аппаратом теории относительности, введённым Минковским в 1908 году \cite{min}. В четырёхмерном мире (4-мире) Минковского временная координата события $ct$ и 3-вектор его пространственного положения $\bf r$ образуют четырёхмерный вектор. 
Аналогичным образом энергия свободного (изолированного) тела $E$ (более точно, $E/c^2$) и три компоненты его импульса $\bf p$ (более точно, ${\bf p}/c$) образуют четыре компоненты псевдоевклидова 4-вектора. Масса тела $m$ представляет собой скаляр, равный длине этого 4-вектора: 
\begin{equation}
m^2=E^2c^{-4}-{\bf p}^2c^{-2}. 
\end{equation}
(Слово "`псевдоевклидовый"' указывает на то, что квадрат длины 4-вектора равен не сумме, а разности квадратов его $E$- и {$\bf p$}- компонент.)

Тейлор и Уилер в книге \cite{tw} откладывают величины энергии  и импульса по ортогональным осям, а на гипотенузе прямоугольного треугольника откладывают величину массы с помощью утолщённого отрезка.  Но можно просто изобразить соотношение (5) в виде прямоугольного треугольника, если переписать его в виде $E^2=m^2c^4+{\bf p}^2c^2$ (смотрите статью \cite{lo5}). Тогда энергия -- гипотенуза, а масса и импульс -- катеты. Отсюда, в частности, следует, что 
\begin{equation}
E_K=\sqrt{m^2c^4+{\bf p}^2c^2}-mc^2.
\end{equation}
 
Основное соотношение специальной теории относительности (5) проверено в тысячах опытов с точностью до десятка значащих цифр. Из него, для тела, обладающего массой, при импульсе, равном нулю, следует формула $E_0=mc^2$, где индекс ноль у энергии
указывает на то, что $E_0$ это энергия не движущегося, а именно покоящегося тела. Если же переписать это же соотношение в виде $(Ec^{-2} - m)(Ec^{-2} + m)={\bf p}^2c^{-2}$, то при $E_K \ll E_0$ без разложения квадратного корня получается Ньютоновское $E_K={\bf p}^2/2m$. 
Аналогичным образом для релятивистских частиц получаем $E-|{\bf p}|c=mc^2/2E$. (Это уравнение очень существенно при описании нейтринных осцилляций.)  
Из того же соотношения (5) и формулы (4) для скорости
${\bf v}={\bf p}c^2/E $
следует, что скорость безмассового фотона при любой энергии равна $c$. 

Так что специальная теория относительности безупречна, чего нельзя сказать о её популяризациях. 

К сожалению, внезапная болезнь и смерть Минковского не позволили ему убедить его современников в преимуществах четырёхмерного мира, и они продолжали "`засовывать"' содержание теории относительности в трёхмерные уравнения механики Ньютона. Хотя Эйнштейн использовал четырёхмерие при формулировке общей теории относительности как теории гравитационного взаимодействия, уравнения $E^2c^{-4}-{\bf p}^2c^{-2}=m^2$ в его собрании сочинений мне найти не удалось. Впервые оно появилось в статьях Клейна \cite{kle}, Фока \cite{fock}, Гордона \cite{gor} (1926) и особенно трудах Дирака \cite{dir} (1930), в которых была создана релятивистская квантовая механика. (Эйнштейн, несмотря на то, что он стоял у истоков понятия кванта, так и не принял квантовой механики.) И только позднее эта формула появилась не только в квантовой, но и в классической электродинамике, в учебнике "`Теория поля"' Ландау и Лифшица в 1941 году (на русском языке) \cite{ll}. 

Четырёхмерие позволило единым образом описывать как массивные, так и безмассовые частицы материи. Оно позволило понять, что масса и материя -- это не одно и то же, что энергия и импульс -- это меры всех процессов и всех движений в природе. Что же касается массы частиц, то она становится несущественной в процессах при очень высоких энергиях $E \gg mc^2$.
\section{Скорость света как единица скорости}
Существование универсальной максимальной скорости $c$ даёт возможность выражать любую скорость $v$ в единицах $c$ как безразмерное число $\beta=v/c$. При этом очевидно, что для самой $c$ получается $\beta=1$. Это позволяет вообще избавиться от буквы $c$  в уравнениях теории относительности, записав уравнения (2), (4), (6) в виде
\begin{align}
E_0=m,\hspace{7mm}
m^2=E^2-{\bf p}^2,\hspace{7mm}
{\bf v}={\bf p}/E.
\end{align}
Уравнение же (3) $E=mc^2$ сводится к $E=m$. Оно явно противоречит уравнению (1) $E=E_K+E_0=E_K+m$  и, следовательно, неправильно.

\section{Заключение}
В 1980-х годах Володя Грибов, взгляды которого на $E=mc^2$ полностью совпадали с моими, по-дружески советовал мне не бороться с этим неверным уравнением, т.к. победить его нельзя. C ощущением постоянного поражения писал я 
предыдущий текст для тома "`Грибов-80"' летом 2010 года во время небывалой жары и смога в Москве, что не могло  не ухудшить качество текста. 
Осенью, несколько недель тому назад, Юлия Нири напомнила мне, что этот текст является по существу продолжением того, что я опубликовал в томе "`Грибов-75"' \cite{lo7}. Там я сравнил формулу $E=mc^2$ с вирусом. Действительно, понятие релятивистской массы,запрятанное в этой формуле, -- это семантический вирус, очень похожий на компьютерные вирусы. Заражённые им люди (они часто называют себя релятивистами) убеждены, что Релятивистская Масса -- это основной портал, ведущий в здание Теории Относительности. Они убеждены в этом,
потому что считают массу мерой инерции. Они игнорируют тот факт, что масса является мерой инерции только для очень медленных тел и частиц, у которых энергия покоя $E_0$ настолько больше, чем кинетическая энергия $E_K$, что кинетической энергией можно пренебречь. При не очень малых скоростях масса -- это только приближённая мера инерции. Для быстрых же частиц (фотонов, нейтрино, протонов в Большом Адронном Коллайдере), для которых $E_K \gg mc^2$, мерой инерции является их энергия $E$.

Общеизвестно, что формулы в физике -- это продолжение обычного языка: уравнения -- это закодированные фразы, а символы, из которых состоят уравнения, заменяют обычные слова или термины. Чтобы не возникала путаница, каждый символ должен быть однозначно связан с соответствующим ему термином. Можно ли ввести в теорию относительности  понятие релятивистской массы? Безусловно -- можно: $m_r=E/c^2$, хотя это будет просто другим обозначением энергии, поскольку $c$ -- универсальная константа. Можно ли вслед за этим ввести понятие релятивистской массы покоя: $m_{r0}=m$? Безусловно -- можно. Конечно, можно ввести и то, и другое; но незачем, потому что СТО и без $m_r$, и без $m_{r0}$ является законченной, самосогласованной  теорией. А вот эту релятивистскую массу покоя  $m_{r0}$ обозначать символом $m_0$ и называть её просто массой покоя уже нехорошо, т. к. это уже как бы подразумевает, что в СТО масса $m$ зависит от скорости. А все мы прекрасно знаем, что в СТО $m$ -- лоренцовский инвариант: масса $m$ одна и та же в покое и в движении, и навешивать на неё индексы бессмысленно. 

Но настоящая беда возникает, когда $m_r$ называют массой и обозначают символом $m$, а обычную ньютоновскую массу $m$ в СТО называют массой покоя и начинают обозначать  $m_0$. Это смешение двух языков ("`французского с нижегородским"') уродует прекрасную теорию,  приводит к невообразимой путанице, длящейся уже сто лет, препятствует пониманию сути СТО. А "`релятивисты-философы"'  утверждают при этом, что ньютоновская механика якобы не является предельным случаем эйнштейновской и что эти две теории якобы несоизмеримы. 
              
В результате обратного воздействия массовой культуры на научную литературу, в лучшем учебнике по физике ХХ века, вышедшем в 1960-х годах, -- "`Фейнмановских лекциях по физике"'\cite{fey} во многих главах повторяется утверждение, что согласно теории относительности, масса растёт с ростом скорости. То же утверждает и маленькая популярная книжечка Ландау и Румера "`Что такое теория относительности?"' \cite{lr}. (Эта книга была написана в 1930-х годах, до ареста обоих авторов, а вышла из печати в 1950-х годах, после освобождения Румера из ссылки. (Ландау пробыл в тюрьме год.))
"`Теория поля"' Ландау и Лифшица \cite{ll}, вышедшая из печати в 1940-х годах, была первым учебником в мире, в котором масса последовательно не зависела от скорости. Тем не менее, понятие энергии покоя и символ $E_0$ в ней отсутствовали, а формула Эйнштейна была дана в форме $E=mc^2$. Это несоответствие сохранилось даже в последнем издании книги, вышедшем в ХХI веке. Поистине, никто не совершенен. Не совершенен наш язык: "`мысль изречённая есть ложь"'.

Невозможно  в этой краткой заметке перечислить книги и статьи авторов, создавших теорию относительности, но их легко найти, если кликнуть гипер-ссылки \cite{lo3}, \cite{lo4}, \cite {lo5},\cite{lo7}, \cite{lo1}, \cite{lo2}, приведённые ниже. Седьмая гипер-ссылка \cite{lo6} содержит слайды доклада о том, что всё преподавание физики 
должно быть основано на двух фундаментальных константах природы: $c$ и $\hbar$. Оперируя этими двумя константами, я предполагаю изложить основы физики в маленькой книжечке "`Азы физики"' на ста страницах.

\vspace{3mm}

{\bf Благодарности}

\vspace{3mm}

Я благодарен Эрике Гуляевой, Мареку Карлинеру, Эле и Виталию Кисиным, Дмитрию Надёжину, Борису Окуню и Зурабу Силагадзе, общение с ними помогло мне написать эту заметку.  

Работа поддержана грантом Президента РФ  НШ-4172.2010.2.